 \documentclass[final,5p,times,twocolumn,authoryear]{elsarticle}

\usepackage{amssymb}
\usepackage{amsmath}
\usepackage{lipsum}
\usepackage{caption}
\usepackage{multirow}
\usepackage{xcolor}
\usepackage{url}
\newcommand{\kms}{\ensuremath{\mathrm{km\,s^{-1}}}}
\newcommand{\farcm}{\mbox{$.\mkern-4mu^\prime$}}

\DeclareUnicodeCharacter{2212}{\textminus}

\journal{New Astronomy}

\begin{document}

\begin{frontmatter}



\title{Modelling the 6.7 GHz Methanol Flaring in G358.93-0.03 Using the Maxwell-Bloch Framework}


\author[uwo]{Ningyan Fang\corref{cor1}}\ead{nfang6@uwo.ca}
\author[mac]{Toktam Rashidi}
\author[Ibaraki]{Yoshinori Yonekura}
\author[uwo,mac]{Vahid Anari}
\author[Ibaraki,RIKEN]{Ross A. Burns}
\author[uwo]{Martin Houde}
\author[mac]{Fereshteh Rajabi}
\cortext[cor1]{Corresponding author}

\affiliation[uwo]{organization={Department of Physics and Astronomy, The University of Western Ontario},
            addressline={1151 Richmond Street}, 
            city={London},
            postcode={N6A 3K7}, 
            state={Ontario},
            country={Canada}}
            
\affiliation[mac]{organization={Department of Physics and Astronomy, McMaster University},
            addressline={1280 Main Street West}, 
            city={Hamilton},
            postcode={L8S 4L8}, 
            state={Ontario},
            country={Canada}}
            
%\affiliation[Yama]{organization={Research Institute for Time Studies, Yamaguchi University},
            %addressline={1677-1 Yoshida},
            %city={Yamaguchi},
            %postcode={753-8511},
            %state={Yamaguchi},
            %country={Japan}}

\affiliation[Ibaraki]{organization={Center for Astronomy, Ibaraki University},
            addressline={2-1-1 Bunkyo},
            city={Mito},
            postcode={310-8512},
            state={Ibaraki},
            country={Japan}}
\affiliation[RIKEN]{organization={RIKEN Pioneering Research Institute},
            addressline={2-1 Hirosawa},
            city={Wako-shi},
            postcode={351-0198},
            state={Saitama},
            country={Japan}}

\begin{abstract}

We model selected 6.7~GHz methanol flare profiles in
G358.93$-$0.03 using the one-dimensional Maxwell--Bloch framework.
The 2019 non-periodic flare, associated with an accretion burst,
produced complex single-dish light curves characterized by sharp rises,
multiple peaks, secondary structure, and extended decays. For six
representative velocity channels, we model each light curve as a sum of
scaled intensity responses from effective emitting systems driven by
phenomenological pump pulses of fixed width but with
component-dependent background rates, amplitudes, and arrival times.
The inferred initial inverted column densities correspond to
characteristic superradiance timescales
$T_{R,0} \simeq 0.3$--$1.4~{\rm d}$, which are shorter than the adopted
dephasing and relaxation timescales, $T_2 = 6.9~{\rm d}$ and
$T_1 = 144.4~{\rm d}$. The fitted components therefore lie in the
fast-transient superradiance regime. Existent VLBI observations motivate the
possibility that a single-dish velocity channel blends emission from
multiple unresolved flaring regions, although these observations are
not used as direct fitting constraints. Although the decomposition is
not unique, it provides a physically motivated description of the
complex flare morphology within a unified Maxwell--Bloch framework.

\end{abstract}



\end{frontmatter}
%\tableofcontents



%%%%%%%%%%%%%%%%%%%%%%%%%%%%%%%%%%%%%%%%%%%%%%%%%%%%%%%%%%%%%%%%%%%%%%%%%%%%%
%%%%%%%%%%%%%%%%%%%%%%%%%%%%%%%%%%%%%%%%%%%%%%%%%%%%%%%%%%%%%%%%%%%%%%%%%%%%%

\section{Introduction}
\label{sec:introduction}

Class II methanol masers are closely associated with high-mass young stellar objects and are radiatively pumped by infrared emission from warm dust \citep{Sobolev1997,Cragg2005,Breen2013}. The 6.7~GHz transition is among the most widely monitored maser lines in massive star-forming regions, and its variability provides a sensitive probe of changes in the local excitation environment \citep{Goedhart2004,VanDerWalt2009,Fujisawa2014,Szymczak2018a,Szymczak2018b}. Large non-periodic flares are especially useful in this context, since they can be associated with sudden changes in the radiation field produced by episodic accretion, for example.

G358.93$-$0.03 (hereafter G358) is a well-studied example of a non-periodic 6.7~GHz methanol flare associated with an accretion burst. The source is a massive star-forming region at a kinematic distance of $6.75^{+0.37}_{-0.68}$~kpc \citep{Brogan2019}. In January 2019, the Ibaraki Maser Monitoring program (iMet) detected the onset of a strong 6.7~GHz Class II methanol flare \citep{Sugiyama2019}. The flux density increased from a previously reported level of $\sim 5$~Jy \citep{Caswell2010} to 99~Jy within two weeks, and later reached a peak of order $10^3$~Jy within about two months \citep{MacLeod2019}. Infrared observations confirmed that the event was associated with an accretion burst from a massive young stellar object \citep{Stecklum2021}.

The same event was accompanied by rich maser activity, including the detection and monitoring of several methanol transitions and rare torsionally excited lines \citep{MacLeod2019,Miao2022,Zhang2024,Johnson2025}. Multi-epoch VLBI observations of the 6.7~GHz methanol masers were used by \citet{Burns2020} to trace the propagation of a thermal-radiation heat wave through the masing region, while later observations were interpreted by \citet{Burns2023} as evidence for a Keplerian disk with a four-arm spiral structure around G358.93$-$0.03-MM1. Based on previous Atacama Large Millimeter/submillimeter Array (ALMA) observations, G358.93$-$0.03 has been resolved into a protocluster of eight continuum sources, among which MM1 and MM3 are identified as molecular hot cores \citep{Brogan2019}. The brightest source MM1 exhibits rich spectra and is associated with maser activities and the accretion burst \citep{Brogan2019,Breen2019,Bayandina2022}. It is therefore highly relevant to this work. These results indicate that the maser emission arises from a spatially structured and evolving environment. Since the light curves analysed here are single-dish measurements, a given velocity channel will include unresolved contributions from more than one masing region.

Although the accretion burst provides a plausible source of enhanced infrared excitation, it does not by itself determine the detailed temporal morphology of the 6.7~GHz flare. The observed single-dish light curves are not simple isolated pulses: neighbouring velocity channels show sharp rises, multiple peaks, secondary structures, and extended decays. The main question addressed in this work is therefore how the flaring medium responds to the excitation associated with the 2019 event, rather than what triggered the accretion burst itself.

Dicke's superradiance has previously been proposed as a fast transient radiation mechanism in maser-hosting regions, where population inversion and velocity coherence are already present \citep{Rajabi2016,Rajabi2017,Rajabi2020}. Earlier applications included 6.7~GHz methanol and 22~GHz water flares in star-forming regions \citep{Rajabi2017}, OH 1612 MHz in circumstellar envelopes of evolved stars \citep{Rajabi2016}, followed by non-periodic 6.7~GHz methanol flares in S255IR-NIRS3, G24.329$+$0.144, and Cep~A \citep{Rajabi2019}. The Maxwell--Bloch framework was then used to clarify the connection between the quasi-steady maser regime and the fast transient superradiance regime \citep{Rajabi2020}, and has since been applied to periodic and multi-transition flares in sources such as G9.62$+$0.20E \citep{Rajabi2023,Rashidi2026} and G22.356$+$0.066 \citep{Rashidi2025}.

Here we apply this framework to selected 6.7~GHz velocity components of
the 2019 flare in G358. For each velocity channel, we model the
single-dish light curve as a sum of scaled intensity responses from
multiple effective one-dimensional Maxwell--Bloch emitting systems,
each driven by a phenomenological pump function representing its local
excitation history. This construction is motivated by VLBI observations
suggesting that a single-dish channel blends emission from several
unresolved flaring regions, although these observations are not used as
direct fitting constraints. We test whether the complex flare
morphology can be understood as the combined intensity response of
multiple effective emitting systems to a changing excitation
environment.

This paper is organized as follows. Section~\ref{sec:model_framework} summarizes the Maxwell--Bloch framework and model setup. Section~\ref{sec:data} describes the single-dish monitoring data. Section~\ref{sec:results_discussion} presents and discusses the fits to the selected velocity components. Section~\ref{sec:conclusion} gives our conclusions. Finally, the Maxwell--Bloch equations are detailed in \ref{sec:appendix_mbe}, while \ref{app:fit_parameters} shows an example of a fitted multi-component profile with a summary of the fitted parameters for all cases analysed in the paper.

%%%%%%%%%%%%%%%%%%%%%%%%%%%%%%%%%%%%%%%%%%%%%%%%%%%%%%%%%%%%%%%%%%%%%%%%%%%%%

\section{Maxwell--Bloch framework and model setup}
\label{sec:model_framework}

The Maxwell--Bloch equations (MBEs; see \ref{sec:appendix_mbe}) provide a framework for modelling the interaction between an inverted molecular ensemble and its radiation field \citep{MacGillivray1976,Gross1982,Benedict1996}. A population-inverted gas described by the MBEs can evolve in two complementary limits: the quasi-steady maser regime, in which the radiation is sustained by stimulated emission, and the fast transient superradiance regime, in which the molecular ensemble radiates through cooperative spontaneous emission \citep{Feld1980,Rajabi2020}. As previously stated, this framework has been successfully applied to non-periodic 6.7~GHz methanol flares \citep{Rajabi2017,Rajabi2019}, to multi-transition periodic flaring in G9.62$+$0.20E \citep{Rajabi2023}, and to periodic or multi-periodic 6.7~GHz methanol flares in later applications \citep{Rashidi2025,Rashidi2026}.
% Replace Rashidi2025,Rashidi2026 with the exact bibliography keys used in the .bib file,
% e.g. Rashidi2025a,Rashidi2025b, if both Toktam papers have the same publication year.

In the quasi-steady maser regime, the evolution time of the population inversion, polarization, and radiation field is long compared with the non-coherent relaxation and dephasing (e.g., from collisions) timescales, i.e., \(T_e\gg T_1,T_2\). The system can then adjust to slow changes in the excitation, and the emitted intensity can approximately follow the pump or seed-radiation variation. In the fast transient superradiance regime, the evolution time is instead comparable to the cooperative radiation timescale, \(T_e\sim T_{\rm R}\), with \(T_{\rm R}<T_1,T_2\). Coherence can then build up before relaxation and dephasing destroy it, allowing the inverted molecular sample to release its stored energy cooperatively as a short, intense burst of radiation \citep{Rajabi2023, Houde2024}.

This difference in response is central to the present analysis. In the saturated maser limit, the emitted intensity scales approximately linearly with the number of participating molecules, \(N\), whereas in the superradiance regime the peak intensity scales as \(N^2\). The superradiant flare is therefore a dynamical response of the medium and does not need to reproduce the detailed shape of the excitation. For a transition with wavelength \(\lambda\) and Einstein coefficient for spontaneous emission \(\Gamma\), the characteristic superradiance timescale is
\begin{align}
    T_{\rm R} =
    \frac{8\pi}{3nL\lambda^2\Gamma},
    \label{eq:T_R}
\end{align}
where \(n\) is the population inversion density and \(L\) is the effective sample length. For a given transition, the inverted column density \(nL\) controls the cooperative response timescale. The competing timescales \(T_1\) and \(T_2\) are treated here as effective non-coherent relaxation and polarization-dephasing timescales within the two-level MBE description (see equations \ref{eq:mbe_inversion}--\ref{eq:mbe_polarization}).

We solve the one-dimensional MBEs numerically using a fourth-order Runge--Kutta method, following the approach of \citet{Rajabi2019,Rajabi2023} and \citet{Rashidi2025,Rashidi2026}. Each emitting region is approximated as a cylindrical system containing methanol molecules modelled as two-level systems interacting with their common radiation field. The radiation propagates along the symmetry axis of the cylinder, and the temporal evolution is followed in the retarded time, \(\tau=t-z/c\). To minimize diffraction and transverse effects not included in the one-dimensional treatment, each system is assigned a Fresnel number of unity,
\begin{align}
    F=\frac{A}{\lambda L}=1,
    \label{eq:fresnel}
\end{align}
where \(A\) is the cross-sectional area of the cylinder.

The iMet light curves analysed in this work were obtained with the Hitachi 32-m single-dish telescope (see Sec.~\ref{sec:data}). At the adopted distance of G358 (\(d\approx 6.75\)~kpc), the \(4\farcm6\) beam at 6.7~GHz corresponds to a projected linear scale of approximately \(10^6\)~au. A given velocity channel will therefore blend emission from multiple spatially unresolved flaring regions within the telescope beam (as will be seen in Figure \ref{fig:g358_vlbi}, the size of the emitting region is on the order of \(10^3\)~au). We accordingly represent the observed flux density in velocity channel \(v\) as a sum of scaled intensity responses from effective emitting systems described by the one-dimensional
MBEs,
\begin{align}
    S_v(t)=\sum_{j=1}^{N_v} C_{v,j} I_{v,j}(t),
    \label{eq:superposition}
\end{align}
where \(I_{v,j}(t)\) is the intensity response of the \(j\)-th
effective emitting system, \(C_{v,j}\) scales the model intensity to the observed flux-density scale, and \(N_v\) is the number of components used for velocity channel \(v\).
The index \(j\) labels the fitted components, denoted profile~1, profile~2, and so forth in the figures below. Each effective system is driven by its own pump history; the pump functions are not combined. Only the resulting intensity responses are scaled and summed to construct the model single-dish light curve.

The decomposition in equation~(\ref{eq:superposition}) is phenomenological and likely not unique. The fitted components should therefore be interpreted as representative effective responses, rather than as a unique reconstruction of the physical source structure.

The inversion pump used in the calculations is also phenomenological. It represents the local effective excitation experienced by each emitting system, rather than a direct radiative-transfer model of the infrared burst, accretion event, or disk structure. Since the G358 flare is a single non-periodic event, for most components we use a single pump pulse of the form
\begin{align}
    \Lambda_{N,j}(\tau)
    =
    \Lambda_{0,j}
    +
    \frac{\Lambda_{1,j}}
    {\cosh^2[(\tau-\tau_{0,j})/T_{\rm p}]},
    \label{eq:pump}
\end{align}
where \(\Lambda_{0,j}\) is the constant background pump, \(\Lambda_{1,j}\) is the pump-pulse amplitude responsible for the flaring episode, \(T_{\rm p}\) is its pump pulse duration, and \(\tau_{0,j}\) is the arrival time of the excitation for the \(j\)-th sample. The constant component sets the initial inversion level through
\begin{align}
    n_{0,j}=\Lambda_{0,j}T_{1}.
    \label{eq:n0}
\end{align}

For components requiring an extended decay, we use an asymmetric effective pump,
\begin{align}
\Lambda_{N,j}(\tau)=
\begin{cases}
\Lambda_{0,j}
+
\dfrac{\Lambda_{1,j}}
{\cosh^2[(\tau-\tau_{0,j})/T_{\rm p}]},
& \tau \leq \tau_{0,j},\\[1.2ex]
\Lambda_{0,j}+
\dfrac{\Lambda_{1,j}}
{\cosh^2[(\tau-\tau_{0,j})/T_{\rm tail}]},
& \tau > \tau_{0,j},
\end{cases}
\label{eq:pump_tail}
\end{align}
with \(T_{\rm tail}=500\,T_{\rm p}\). This prescription is identical to equation~\eqref{eq:pump} on the rising side, but decays slowly after the pump maximum. The long-tail pump is used only as an effective representation of slowly varying excitation responsible for extended decays. It should not be interpreted as a unique physical light curve of the accretion burst.

The numerical values of \(T_1\), \(T_2\), \(T_{\rm p}\), \(L\), and the pump parameters used in the fits are given in Section~\ref{sec:results_discussion}. The construction of the multi-component fits is illustrated in \ref{app:fit_parameters}, where the effective pumps, population inversion responses, individual intensity responses, and final composite fit are shown together for one representative velocity channel.

%%%%%%%%%%%%%%%%%%%%%%%%%%%%%%%%%%%%%%%%%%%%%%%%%%%%%%%%%%%%%%%%%%%%%%%%%%%%%
\section{Data}
\label{sec:data}

The 6.7~GHz methanol light curves analysed in
this work were obtained from the Ibaraki 6.7~GHz Methanol Maser
Monitoring Database\footnote{\url{https://vlbi.sci.ibaraki.ac.jp/iMet/}}.
As previously stated, we use observations of G358 obtained with the Hitachi
32-m single-dish telescope during the 2019 flaring episode and its
subsequent decay. At 6.7~GHz, the telescope has a beam full width
at half maximum (FWHM) of \(4\farcm6\) \citep{Yonekura2016, Sugiyama2017}, which covers the whole G358 cluster ($\sim$1.5 arcseconds at 6.75~kpc with size up to 10000~au) \citep{Brogan2019}.
The analysed interval extends from approximately MJD~58500 to
MJD~59000, corresponding roughly to 2019 January through 2020 May.
This interval covers the main rise, peak, and decay of the
6.7~GHz flare.

The spectra span the velocity range from
\(-21.22~\kms\) to \(-13.90~\kms\). Before smoothing, the
Hitachi 6.7~GHz observing setup has a velocity resolution of
\(0.044~\kms\). Three-channel smoothing yields an effective
velocity resolution of approximately \(0.13~\kms\) and a
typical \(1\sigma\) rms noise level of about
\(0.3~\mathrm{Jy}\) for a 5-min integration
\citep{Sugiyama2017}. Within each of the two velocity groups
examined here, the selected velocity samples are separated by
\(0.10\)--\(0.12~\kms\), comparable to the effective spectral
resolution, and therefore do not constitute fully independent
spectral resolution elements.

The VLBI maps of the 6.7~GHz methanol masers in G358 \citep{Burns2020,Burns2023} are used in
Section~\ref{sec:results_discussion} to provide spatial context and to show that a single-dish velocity channel will include emission from more than one unresolved emitting region.

%%%%%%%%%%%%%%%%%%%%%%%%%%%%%%%%%%%%%%%%%%%%%%%%%%%%%%%%%%%%%%%%%%%%%%%%%%%%%

\section{Results and discussion}
\label{sec:results_discussion}

\begin{figure}[t!]
    \centering
    \includegraphics[width=\linewidth]{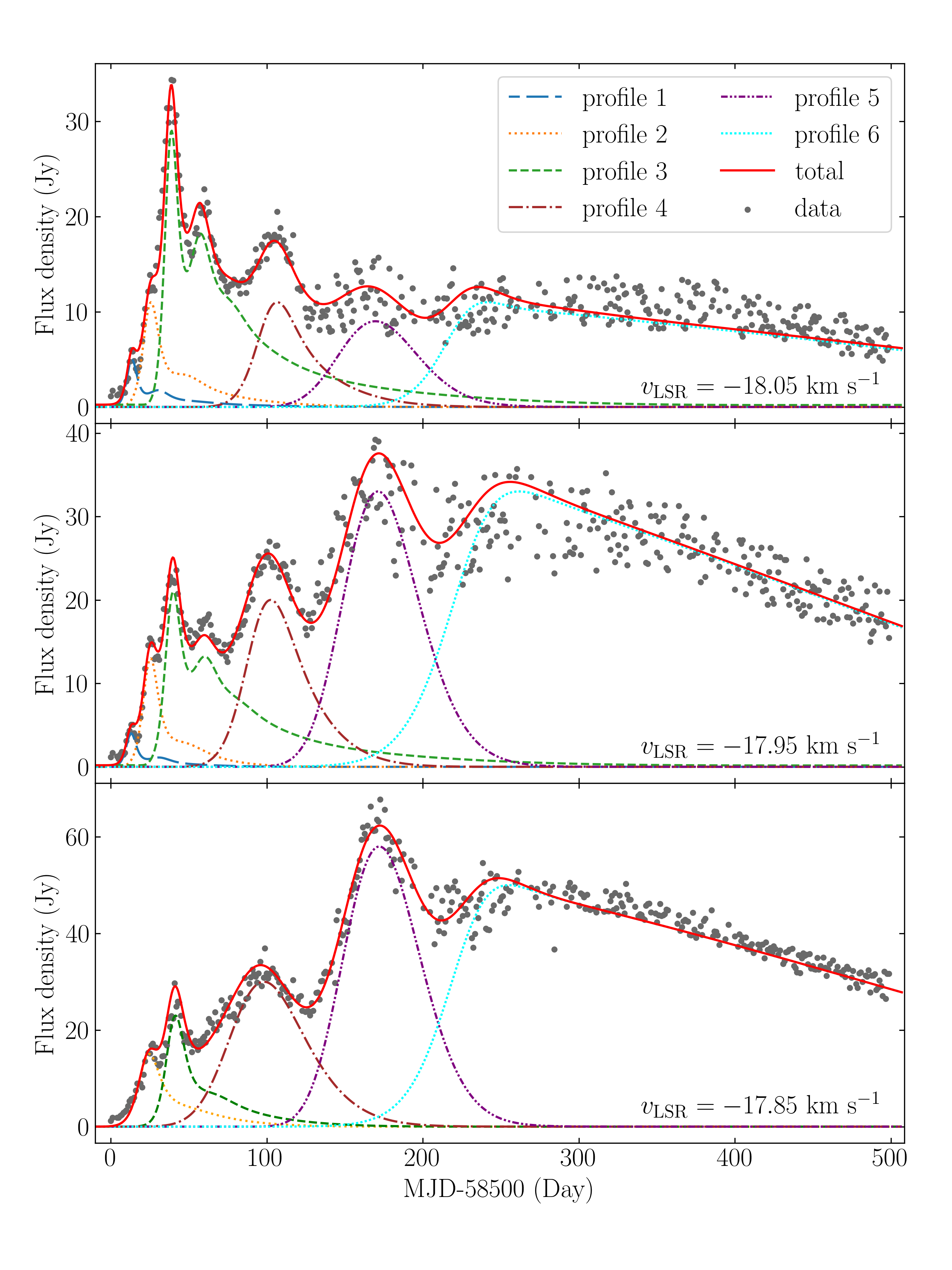}
    \caption{Flares of the 6.7~GHz methanol transition in G358 at
    \(v_{\rm LSR}=-18.05\) (top), \(-17.95\) (middle), and \(-17.85~\kms\) (bottom). Grey points show the iMet single-dish
    light curve, coloured curves show the individual scaled Maxwell--Bloch
    responses, and the solid red curve their sum. The fitted pump
    parameters are listed in Table~\ref{tab:fit_parameters}.}
    \label{fig:v~-18}
\end{figure}

\begin{figure}[t!]
    \centering
    \includegraphics[width=\linewidth]{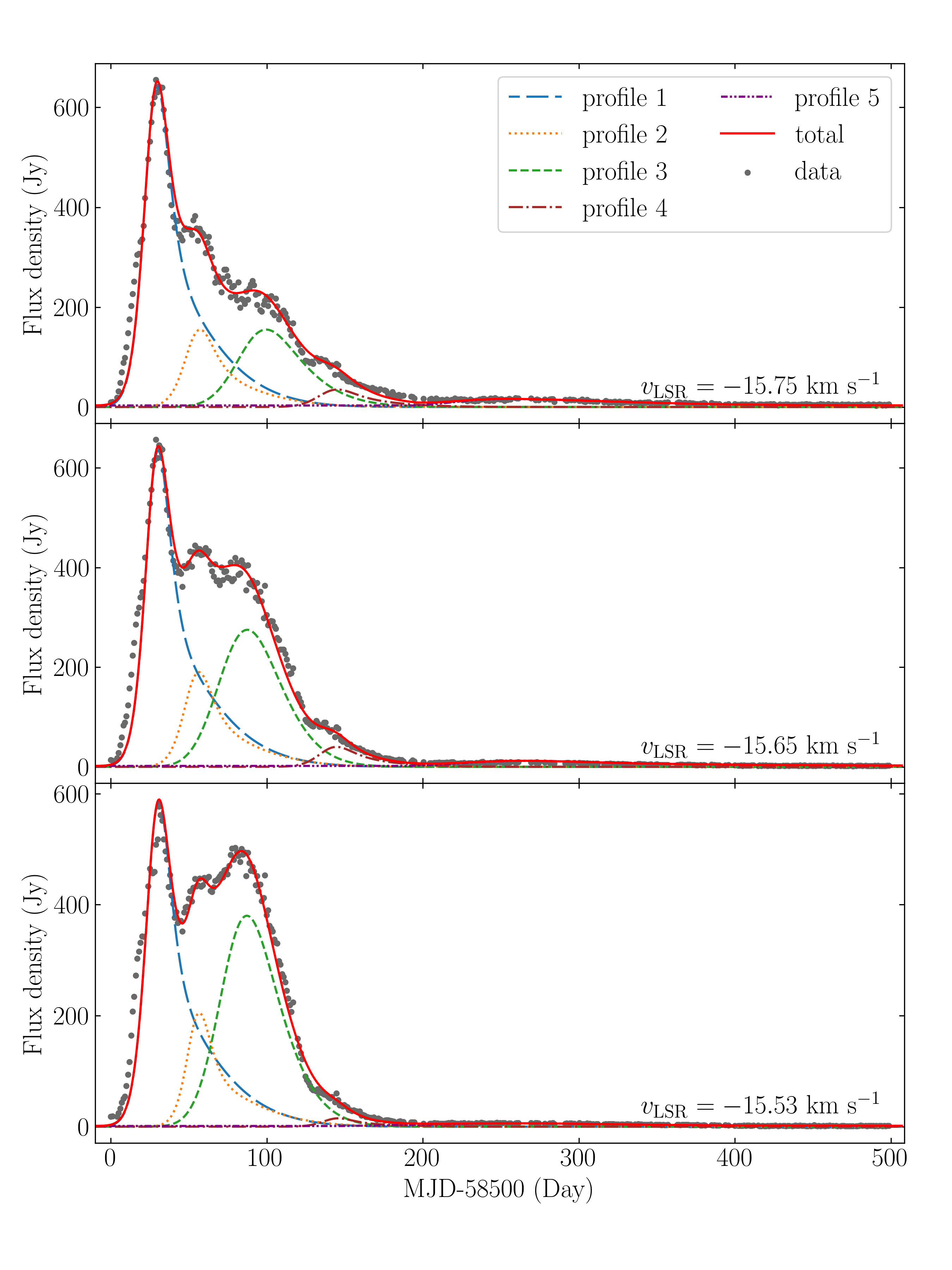}
    \caption{Same as Figure \ref{fig:v~-18} but for \(v_{\rm LSR}=-15.75\) (top), \(-15.65\) (middle), and \(-15.53~\kms\) (bottom).}
    \label{fig:v~-15.6}
\end{figure}

We model the flare profiles in six representative velocity channels:
\(v_{\rm LSR}=-18.05~\kms\), \(-17.95~\kms\), \(-17.85~\kms\), \(-15.75~\kms\), \(-15.65~\kms\) and
\(-15.53~\kms\). These channels were well sampled during the 2019 flare and neighbouring channels show related temporal structure. The channels are discussed in two groups: the \(-18.05~\kms\) to \(-17.85~\kms\) group, shown in Fig.~\ref{fig:v~-18}, and the \(-15.75~\kms\) to \(-15.53~\kms\) group in Fig.~\ref{fig:v~-15.6}.

For all fits, the environmental response parameters are kept fixed at \(T_1=144.4\)~d, \(T_2=6.9\)~d, \(T_{\rm p}=1.7\)~d, and \(L=2.51\times10^{15}\)~cm.  The effective pump amplitudes, arrival times, and scaling coefficients vary among components. The fitted pump parameters are listed in Table~\ref{tab:fit_parameters}. Fluctuations in $T_1$ and $T_2$ due to the spatial distribution of the methanol emission region can be absorbed by changes in the pumps. In Figs.~\ref{fig:v~-18}--\ref{fig:v~-15.6}, the grey points show the iMet single-dish light curves, the coloured dashed and dotted curves show the individual scaled Maxwell--Bloch responses, and the solid red curve in each panel shows their sum. The labels profile~1, profile~2, and so forth refer to individual effective response components rather than separate observed data sets.

%%%%%%%%%%%%%%%%%%%%%%%%%%%
\subsection{Velocity group near \(-18~\kms\)}

The first velocity group, shown in Fig.~\ref{fig:v~-18}, contains a similar sequence of flare components across neighbouring channels. The earliest narrow components reproduce the sharp initial rise, while later components account for the secondary structure and extended decay. A ringing-like secondary structure is reproduced in profile~3 for \(v_{\rm LSR}=-18.05\) and \(-17.95~\kms\), consistent with a fast transient Maxwell--Bloch response. This feature becomes less distinct at \(v_{\rm LSR}=-17.85~\kms\) (bottom panel), where it blends with the neighbouring component. 

The relative strengths of the later components also change across this velocity group. More precisely, profiles~4 and 5 are  more prominent for $v_\mathrm{LSR}=-18.05$ and $-17.95$~\kms than  $v_\mathrm{LSR}=-17.85$~\kms. For all cases, profile 6 shows an initial superradiance transient response to a broader pump component before settling into a saturated maser regime (see \ref{app:fit_parameters}; \citealt{Rajabi2020}). At $v_\mathrm{LSR}=-18.05$~\kms this level is close to the previously reported quiescent level of the source \citep{Caswell2010}, while it is stronger for the other two velocities. This suggests that neighbouring velocity channels are not simply scaled versions of the same flare, but instead contain different mixtures of the same components.

%Also, profile~6 for $v_\mathrm{LSR}=-18.05$~\kms is close to the previously reported quiescent level of the source \citep{Caswell2010}, while it is stronger for the other two velocities. 

%%%%%%%%%%%%%%%%%%%%%%%%%%%
\subsection{Velocity group near \(-15.6~\kms\)}

The second velocity group, shown in Fig.~\ref{fig:v~-15.6}, requires fewer effective components than the \(-18~\kms\) group. Profiles~1 and 2 occur early in the flare and remain relatively unchanged across the group. The central component, i.e., profile 3, becomes increasingly dominant from \(v_{\rm LSR}=-15.75\) to \(-15.53~\kms\), whereas profile 4 remains weak and becomes less important at $v_\mathrm{LSR}=-15.53$~\kms. We also note that the low-intensity profile 5 in this velocity group appears to correspond to profile 6 in Fig.~\ref{fig:v~-18} of the $-18~\kms$ group. As before, these variations are interpreted as changes in the relative contributions of effective components rather than as a unique mapping to individual physical emitting regions.

%%%%%%%%%%%%%%%%%%%%%%%%%%%
\begin{figure*}[th]
    \centering
    \includegraphics[width=\textwidth]{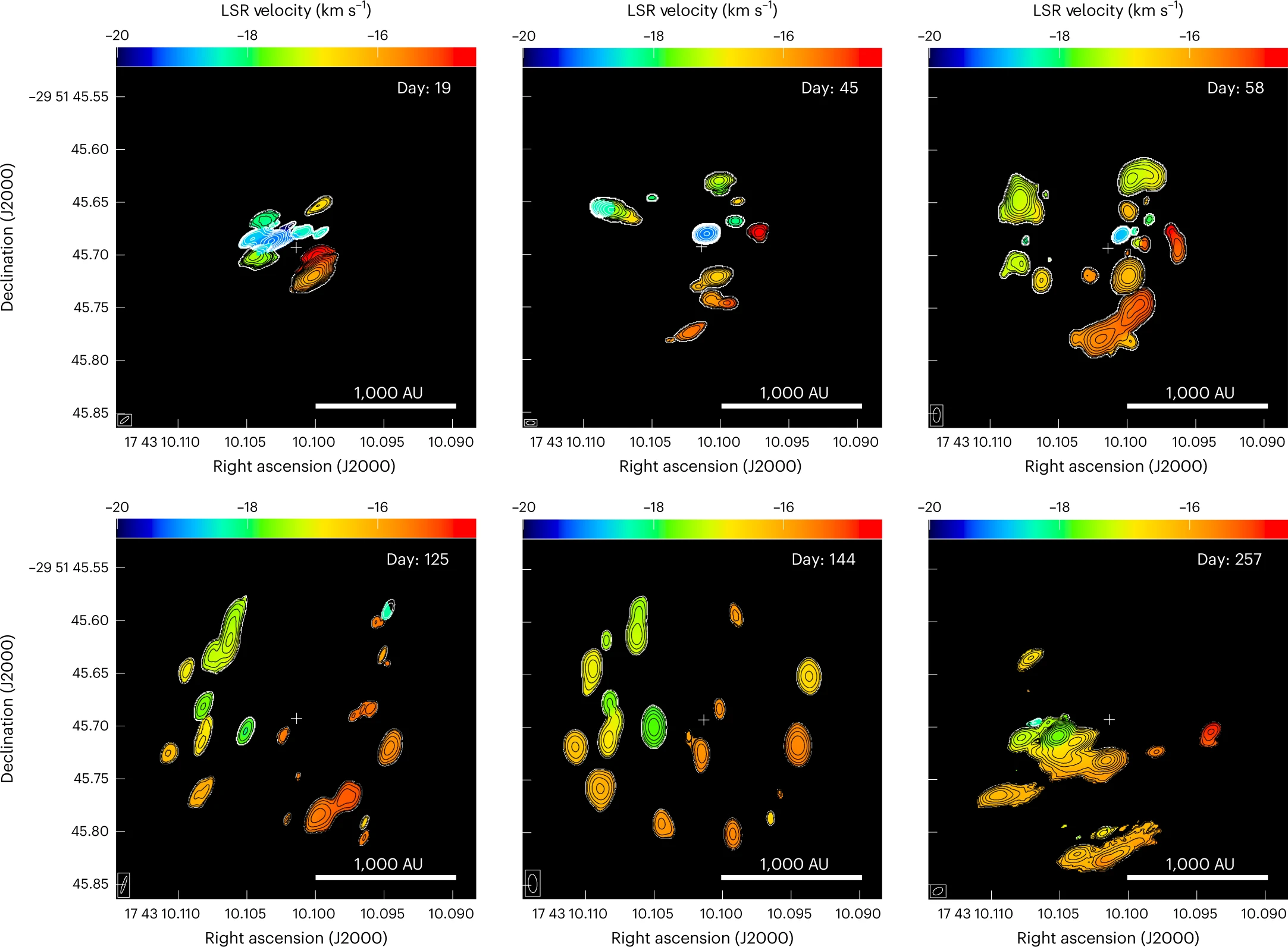}
    \caption{Multi-epoch VLBI maps of the 6.7~GHz methanol masers in G358.93$-$0.03-MM1, reproduced from \citet{Burns2023} with permission. The maps provide spatial context for the single-dish modelling but are not used as fitting constraints. While the linear scale of the emitting region is seen to be on the order of $10^3$~au, that of the single-dish beam is on the order of $10^6$~au. It follows that a single-dish velocity channel blends emission from more than one spatially distinct emitting region.}
    \label{fig:g358_vlbi}
\end{figure*}

\subsection{Timescales and model interpretation}

The fitted background pump rates listed in
Table~\ref{tab:fit_parameters}, together with \(T_1=144.4\)~d and
\(L=2.51\times10^{15}\)~cm, yield initial inverted column densities
of \(n_0L\simeq 10^3\)--\(10^4~\mathrm{cm^{-2}}\). For the 6.7~GHz methanol transition, these values correspond to initial characteristic superradiance timescales of approximately
\(T_{\rm R,0}\simeq 0.3~\mathrm{d}-1.4~\mathrm{d}\). These timescales are shorter than the adopted dephasing and relaxation timescales, \(T_2=6.9\)~d and \(T_1=144.4\)~d. Thus, all fitted components satisfy the basic timescale condition for a fast-transient superradiant response.

The coefficients \(C_{v,j}\) in equation~\eqref{eq:superposition} scale the model intensity of each effective one-dimensional system to the observed single-dish flux density. They absorb uncertainties in the effective emitting solid angle, beaming, and the number of
unresolved samples contributing with similar temporal responses. %This conversion can account for the number of unresolved systems contributing to an intensity component, beaming, geometry, and orientation relative to the line of sight. 

%%%%%%%%%%%%%%%%%%%%%%%%%%%%%%%

\subsection{VLBI context and limitations}

The use of multiple effective systems is motivated by the spatial complexity of G358. The VLBI maps of \citet{Burns2020,Burns2023}, shown here in Fig.~\ref{fig:g358_vlbi}, trace the evolution of the 6.7~GHz methanol emission at several epochs after the flare onset in G358.93$-$0.03-MM1. The emission regions evolve from a compact distribution into a more extended structure, which \citet{Burns2023} interpreted as the propagation of a thermal radiation wave rather than physical motion of the gas. These maps are not used as fitting data, but they establish the fact that a single-dish velocity channel blends emission from more than one spatially distinct emitting region. The fitted Maxwell--Bloch components should therefore not be interpreted as one-to-one identifications of individual VLBI maser spots. We also note that the velocity groups near $-18$ and $-15.6$~\kms form spatially distinct clusters, which could explain the different features and structure seen in the light curves.

The characteristic size of an effective Maxwell--Bloch system is many orders of magnitude smaller than the spatial scales resolved by VLBI. A VLBI maser spot or region may therefore encompass many unresolved effective samples rather than correspond to a single Maxwell--Bloch sample.

The adopted dephasing timescale can be compared with a simple collision-timescale estimate,
\begin{align}
\tau_{\rm coll}\sim
\frac{1}{\sigma_{\rm g}\bar{v}n_{\rm H_2}}.\label{eq:tcoll}
\end{align}
For \(\tau_\mathrm{coll}=T_2=6.9~\rm{d}\), \(\sigma_{\rm g}\sim10^{-16~}\mathrm{cm^2}\),
\(\bar{v}\sim10^5~\mathrm{cm\,s^{-1}}\), equation (\ref{eq:tcoll}) yields
\[
n_{\rm H_2}\simeq1.68\times10^5~\mathrm{cm}^{-3} .
\]
This value lies within, but toward the lower end of, the
\(10^4\)--\(10^9~\mathrm{cm^{-3}}\) density range reported by
\citet{Cragg2005} for the 6.7~GHz methanol transition.

%This falls within the density range of $10^4~\mathrm{cm}^{-3}\leq n_{\mathrm{H}_2}\leq10^9~\mathrm{cm}^{-3}$ for 6.7~GHz methanol line reported by \citet{Cragg2005}. %Given the uncertainty in the local gas density, temperature, and collision rates, \(T_2\) should be regarded as an effective dephasing parameter in the two-level Maxwell--Bloch model.

%%%%%%%%%%%%%%%%%%%%%%%%%%%%%%%%%%%%%%%%%%%%%%%%%%%%%%%%%%%%%%%%%%%%%%%%%%%%%

\section{Conclusion}
\label{sec:conclusion}
We have modelled selected 6.7~GHz methanol flare profiles in G358 using the one-dimensional Maxwell--Bloch framework. The analysed light curves correspond to the 2019 non-periodic flaring episode associated with an accretion burst. Rather than modelling the accretion burst itself, we treated the pump functions phenomenologically as local excitation histories experienced by effective emitting systems.

The selected single-dish profiles can be represented as sums of effective one-dimensional Maxwell--Bloch responses. In these representative fits, the environmental response parameters are kept fixed, while the effective pump amplitudes, arrival times, and scaling coefficients vary between components. This approach allows the complex flare morphology, including sharp rises, secondary structure, and extended decays, to be organized in terms of multiple effective responses of the emitting gas.

The flare components are consistent with fast transient Maxwell--Bloch responses in the superradiance regime. For the current set of fitted parameters, the inferred characteristic superradiance timescales are shorter than the adopted relaxation and dephasing timescales, satisfying the basic condition required for coherent transient emission. Components with slowly decaying tails are reproduced using an effective long-tail pump and should be interpreted as phenomenological representations of slowly varying local excitation, rather than as direct measurements of the accretion-burst luminosity curve.

The use of multiple effective samples within a single velocity channel is motivated by the spatially structured maser environment inferred from VLBI observations of G358.93$-$0.03-MM1. Our multi-component fits are therefore likely not unique, but provide a useful way of describing the complex single-dish flare profiles within a unified Maxwell--Bloch response framework.

Future modelling would benefit from simultaneous high-cadence radio and infrared monitoring, together with spatially resolved observations during the flare evolution. Such data would help constrain the local excitation history more directly and test whether individual flare components can be associated with distinct spatial regions in the emitting environment.

%%%%%%%%%%%%%%%%%%%%%%%%%%%%%%%%%%%%%%%%%%%%%%%%%%%%%%%%%%%%%%%%%%%%%%%%%%%%%

\section*{Acknowledgments}

F.R.’s research is supported by the Natural Sciences and Engineering Research Council of Canada (NSERC) Discovery Grant RGPIN-2024-06346. M.H.’s research is funded through the NSERC Discovery Grant RGPIN-2024-05242. This work is partially supported by the Inter-university collaborative project "Japanese VLBI Network (JVN)" of NAOJ and by the Japan Society for the Promotion of Science (JSPS) KAKENHI Grant Numbers JP21H01120 and JP21H00032.

%%%%%%%%%%%%%%%%%%%%%%%%%%%%%%%%%%%%%%%%%%%%%%%%%%%%%%%%%%%%%%%%%%%%%%%%%%%%%

\section*{Declaration of generative AI and AI-assisted technologies in the manuscript preparation process}

During the preparation of this work, the authors used ChatGPT to improve readability through grammar and language refinement. The authors reviewed and edited the output as needed and take full responsibility for the content of the publication.

%%%%%%%%%%%%%%%%%%%%%%%%%%%%%%%%%%%%%%%%%%%%%%%%%%%%%%%%%%%%%%%%%%%%%%%%%%%%%

\bibliographystyle{elsarticle-harv} 
\bibliography{ref}

@article{Yonekura2016,
  title={The {H}itachi and {T}akahagi 32 m radio telescopes: {U}pgrade of the antennas from satellite communication to radio astronomy},
  author={Yonekura, Yoshinori and others},
  journal={Publications of the Astronomical Society of Japan},
  volume={68},
  number={5},
  pages={74},
  year={2016},
  publisher={Oxford University Press}
}

@article{Burns2020,
   author = {R. A. Burns and others},
   doi = {10.1038/s41550-019-0989-3},
   issn = {23973366},
   issue = {5},
   journal = {Nature Astronomy},
   title = {A heatwave of accretion energy traced by masers in the {G}358-{MM}1 high-mass protostar},
   volume = {4},
   year = {2020}
}

@article{Johnson2025,
   author = {Bradley Johnson and others},
   doi = {10.1017/pasa.2025.10127},
   issn = {14486083},
   journal = {Publications of the Astronomical Society of Australia},
   title = {First detections of methanol maser lines from a rare transition family},
   volume = {42},
   year = {2025}
}

@article{Miao2022,
   author = {Dan Miao and others},
   doi = {10.3847/1538-4365/ac9524},
   issn = {0067-0049},
   issue = {1},
   journal = {The Astrophysical Journal Supplement Series},
   title = {New Methanol Maser Transitions and Maser Variability Identified from an Accretion Burst Source {G}358.93-0.03},
   volume = {263},
   year = {2022}
}

@article{MacLeod2019,
   author = {G. C. MacLeod and others},
   doi = {10.1093/mnras/stz2417},
   issn = {13652966},
   issue = {3},
   journal = {Monthly Notices of the Royal Astronomical Society},
   title = {Detection of new methanol maser transitions associated with {G}358.93-0.03},
   volume = {489},
   year = {2019}
}

@article{Stecklum2021,
   author = {B. Stecklum and others},
   doi = {10.1051/0004-6361/202039645},
   issn = {14320746},
   journal = {Astronomy and Astrophysics},
   title = {Infrared observations of the flaring maser source {G}358.93-0.03: SOFIA confirms an accretion burst from a massive young stellar object},
   volume = {646},
   year = {2021}
}

@article{Zhang2024,
   author = {Shijie Zhang and others},
   doi = {10.3847/1538-3881/ad73bd},
   issn = {0004-6256},
   issue = {5},
   journal = {The Astronomical Journal},
   title = {Variabilities of Methanol Maser and Thermal Molecular Lines from an Accretion Burst Source {G}358.93-0.03},
   volume = {168},
   year = {2024}
}

@article{Bayandina2022,
   author = {O. S. Bayandina and C. L. Brogan and R. A. Burns and X. Chen and T. R. Hunter and S. E. Kurtz and G. C. MacLeod and A. M. Sobolev and K. Sugiyama and I. E. Val’tts and Y. Yonekura},
   doi = {10.3847/1538-3881/ac42d2},
   issn = {0004-6256},
   issue = {2},
   journal = {The Astronomical Journal},
   title = {A Multitransition Methanol Maser Study of the Accretion Burst Source {G358.93–0.03-MM1}},
   volume = {163},
   year = {2022}
}

@article{Caswell2010,
   author = {J. L. Caswell and others},
   doi = {10.1111/j.1365-2966.2010.16339.x},
   issn = {13652966},
   issue = {2},
   journal = {Monthly Notices of the Royal Astronomical Society},
   title = {The 6-{GHz} methanol multibeam maser catalogue - {I}. {G}alactic {C}entre region, longitudes 345° to 6°},
   volume = {404},
   year = {2010}
}

@article{Brogan2019,
   author = {C. L. Brogan and others},
   doi = {10.3847/2041-8213/ab2f8a},
   issn = {2041-8205},
   issue = {2},
   journal = {The Astrophysical Journal Letters},
   title = {Sub-arcsecond (Sub)millimeter Imaging of the Massive Protocluster {G}358.93−0.03: Discovery of 14 New Methanol Maser Lines Associated with a Hot Core},
   volume = {881},
   year = {2019}
}

@article{Burns2023,
   author = {R. A. Burns and others},
   doi = {10.1038/s41550-023-01899-w},
   issn = {23973366},
   issue = {5},
   journal = {Nature Astronomy},
   title = {A Keplerian disk with a four-arm spiral birthing an episodically accreting high-mass protostar},
   volume = {7},
   year = {2023}
}

@article{Breen2013,
   author = {S. L. Breen and others},
   doi = {10.1093/mnras/stt1315},
   issn = {13652966},
   issue = {1},
   journal = {Monthly Notices of the Royal Astronomical Society},
   title = {Confirmation of the exclusive association between 6.7-{GHz} methanol masers and high-mass star formation regions},
   volume = {435},
   year = {2013}
}

@article{Sobolev1997,
   author = {A. M. Sobolev and D. M. Cragg and P. D. Godfrey},
   issn = {00046361},
   issue = {1},
   journal = {Astronomy and Astrophysics},
   month = {8},
   pages = {211-220},
   title = {Pumping of class {II} methanol masers: {II}. The 5\_1-6\_0 {A}+ transition},
   volume = {324},
   year = {1997}
}

@article{Cragg2005,
   author = {D. M. Cragg and A. M. Sobolev and P. D. Godfrey},
   doi = {10.1111/j.1365-2966.2005.09077.x},
   issn = {00358711},
   issue = {2},
   journal = {Monthly Notices of the Royal Astronomical Society},
   title = {Models of class {II} methanol masers based on improved molecular data},
   volume = {360},
   year = {2005}
}

@article{Gross1982,
   author = {M. Gross and S. Haroche},
   doi = {10.1016/0370-1573(82)90102-8},
   issn = {03701573},
   issue = {5},
   journal = {Physics Reports},
   title = {Superradiance: An essay on the theory of collective spontaneous emission},
   volume = {93},
   year = {1982}
}

@book{Benedict1996,
   author = {M G Benedict and A M Ermolaev and V A Malyshev and I V Solokov and E D Trifonov},
   edition = {1st},
   isbn = {9780203737880},
   journal = {Series in Optics and Optoelectronics },
   publisher = {CRC Press},
   title = {Super-radiance: Multiatomic Coherent Emission},
   year = {1996}
}

@article{MacGillivray1976,
   author = {J. C. MacGillivray and M. S. Feld},
   doi = {10.1103/PhysRevA.14.1169},
   issn = {10502947},
   issue = {3},
   journal = {Physical Review A},
   title = {Theory of superradiance in an extended, optically thick medium},
   volume = {14},
   year = {1976}
}

@article{Rajabi2023,
   author = {F. Rajabi and others},
   doi = {10.1093/mnras/stad2671},
   issn = {13652966},
   issue = {1},
   journal = {Monthly Notices of the Royal Astronomical Society},
   title = {Modelling of the multitransition periodic flaring in {G9.62+0.20E}},
   volume = {526},
   year = {2023}
}

@ARTICLE{Rashidi2025,
       author = {{Rashidi}, T. and others},
        title = {Modelling the Periodic 6.7 {GHz} Methanol Flaring in {G}22.356+0.066},
      journal = {Monthly Notices of the Royal Astronomical Society},
         year = {2025},
        month = sep,
       volume = {542},
       number = {1},
        pages = {L12-L18},
          doi = {10.1093/mnrasl/slaf061}
}

@article{Rajabi2020,
   author = {Fereshteh Rajabi and Martin Houde},
   doi = {10.1093/mnras/staa1067},
   issn = {13652966},
   issue = {4},
   journal = {Monthly Notices of the Royal Astronomical Society},
   title = {Astronomical masers and {D}icke's superradiance},
   volume = {494},
   year = {2020}
}

@article{Rajabi2019,
   author = {F. Rajabi and M. Houde and A. Bartkiewicz and M. Olech and M. Szymczak and P. Wolak},
   doi = {10.1093/mnras/stz074},
   issn = {13652966},
   issue = {2},
   journal = {Monthly Notices of the Royal Astronomical Society},
   title = {New evidence for {D}icke's superradiance in the 6.7 {GHz} methanol spectral line in the interstellar medium},
   volume = {484},
   year = {2019}
}

@article{Rajabi2016,
   author = {Fereshteh Rajabi and Martin Houde},
   doi = {10.3847/0004-637x/828/1/57},
   issn = {15384357},
   issue = {1},
   journal = {The Astrophysical Journal},
   title = {DICKE’S SUPERRADIANCE IN ASTROPHYSICS. II. THE {OH} 1612 {MHz} LINE},
   volume = {828},
   year = {2016}
}

@article{Rajabi2017,
   author = {Fereshteh Rajabi and Martin Houde},
   doi = {10.1126/sciadv.1601858},
   issn = {23752548},
   issue = {3},
   journal = {Science Advances},
   title = {Explaining recurring maser flares in the {ISM} through large-scale entangled quantum mechanical states},
   volume = {3},
   year = {2017}
}

@article{Houde2024, 
    title={Variability, flaring and coherence – the complementarity of the maser and superradiance regimes}, 
    volume={18}, 
    DOI={10.1017/S1743921323002764}, 
    number={S380}, 
    journal={Proceedings of the International Astronomical Union}, 
    author={Houde, Martin and others}, 
    year={2024}, 
    pages={399–413}
}

@ARTICLE{Sugiyama2019,
       author = {{Sugiyama}, Koichiro and {Saito}, Yu and {Yonekura}, Yoshinori and {Momose}, Munetake},
        title = "{Bursting activity of the 6.668-{GHz} {CH3OH} maser detected in {G}358.93-00.03 using the {H}itachi 32-m}",
      journal = {The Astronomer's Telegram},
         year = {2019},
        month = jan,
       volume = {12446},
        pages = {1}
}

@article{Rashidi2026,
    author = {Rashidi, T and others},
    title = {A unified Maxwell–Bloch framework for multiperiodic 6.7 {GHz} methanol flaring in {G9.62+0.20E}},
    journal = {Monthly Notices of the Royal Astronomical Society},
    volume = {545},
    number = {3},
    pages = {staf2200},
    year = {2026},
    month = {01},
    issn = {0035-8711},
    doi = {10.1093/mnras/staf2200},
    url = {https://doi.org/10.1093/mnras/staf2200}
}

@ARTICLE{Feld1980,
       author = {{Feld}, M.~S. and {MacGillivray}, J.~C.},
        title = "{Superradiance}",
      journal = {Topics in Current Physics},
         year = 1980,
        month = jan,
       volume = {21},
        pages = {7-57},
          doi = {10.1007/978-3-642-81495-2_2}
}

@article{Sugiyama2017,
   author = {Koichiro Sugiyama and others},
   doi = {10.1093/pasj/psx034},
   issn = {2053051X},
   issue = {4},
   journal = {Publications of the Astronomical Society of Japan},
   title = {The shortest periodic and flaring flux variability of a methanol maser emission at 6.7 {GHz} in {G} 014.23−00.50},
   volume = {69},
   year = {2017}
}

@article{VanDerWalt2009,
   author = {D. J. Van Der Walt and S. Goedhart and M. J. Gaylard},
   doi = {10.1111/j.1365-2966.2009.15147.x},
   issn = {13652966},
   issue = {2},
   journal = {Monthly Notices of the Royal Astronomical Society},
   title = {Periodic class II methanol masers in G9.62+0.20E},
   volume = {398},
   year = {2009}
}

@article{Szymczak2018b,
   author = {M. Szymczak and M. Olech and P. Wolak and E. Gérard and A. Bartkiewicz},
   doi = {10.1051/0004-6361/201833443},
   issn = {14320746},
   journal = {Astronomy and Astrophysics},
   title = {Giant burst of methanol maser in S255IR-NIRS3},
   volume = {617},
   year = {2018}
}

@article{Szymczak2018a,
   author = {M. Szymczak and M. Olech and R. Sarniak and P. Wolak and A. Bartkiewicz},
   doi = {10.1093/MNRAS/STX2693},
   issn = {13652966},
   issue = {1},
   journal = {Monthly Notices of the Royal Astronomical Society},
   title = {Monitoring observations of 6.7GHz methanol masers},
   volume = {474},
   year = {2018}
}

@article{Goedhart2004,
   author = {S. Goedhart and M. J. Gaylard and D. J. Van Der Walt},
   doi = {10.1111/j.1365-2966.2004.08340.x},
   issn = {00358711},
   issue = {2},
   journal = {Monthly Notices of the Royal Astronomical Society},
   title = {Long-term monitoring of 6.7-GHz methanol masers},
   volume = {355},
   year = {2004}
}

@article{Fujisawa2014,
   author = {Kenta Fujisawa and others},
   doi = {10.1093/pasj/psu015},
   issn = {2053051X},
   issue = {2},
   journal = {Publications of the Astronomical Society of Japan},
   title = {Observations of 6.7GHz methanol masers with east-asian VLBI network. I. VLBI images of the first epoch of observations},
   volume = {66},
   year = {2014}
}

@article{Breen2019,
   author = {S. L. Breen and A. M. Sobolev and J. F. Kaczmarek and S. P. Ellingsen and T. P. McCarthy and M. A. Voronkov},
   doi = {10.3847/2041-8213/ab191c},
   issn = {2041-8205},
   issue = {2},
   journal = {The Astrophysical Journal Letters},
   title = {Discovery of Six New Class II Methanol Maser Transitions, Including the Unambiguous Detection of Three Torsionally Excited Lines toward G 358.931–0.030},
   volume = {876},
   year = {2019}
}

\appendix

\section{Maxwell--Bloch equations}
\label{sec:appendix_mbe}

The model is based on the one-dimensional Maxwell--Bloch equations (MBEs) for an ensemble of two-level molecules interacting with a common radiation field \citep{Gross1982,Benedict1996, Rajabi2023}. Under the rotating-wave and the slowly varying envelope approximations, these equations can be written as
\begin{align}
    \frac{\partial n'}{\partial \tau}
    &=
    \frac{i}{\hbar}
    \left(P^+E^+ - P^-E^-\right)
    -\frac{n'}{T_1}
    +\Lambda_N ,
    \label{eq:mbe_inversion}
    \\
    \frac{\partial P^+}{\partial \tau}
    &=
    \frac{2id^2}{\hbar}E^-n'
    -\frac{P^+}{T_2}
    +\Lambda_P ,
    \label{eq:mbe_polarization}
    \\
    \frac{\partial E^+}{\partial z}
    &=
    \frac{i\omega_0}{2\epsilon_0 c}P^- .
    \label{eq:mbe_field}
\end{align}

Here \(\tau=t-z/c\) is the retarded time, where \(c\) denotes the speed of light in vacuum. Furthermore, \(d=|\mathbf{d}|\) is the electric dipole moment of the transition, \(\omega_0\) is the transition angular frequency, and \(\epsilon_0\) is the vacuum permittivity. The total population inversion density is \(n=2n'\), while \(P^+\) and \(E^+\) are the slowly varying amplitudes of the induced polarization and electric field, respectively.

% Here \(\tau=t-z/c\) is the retarded time, \(d=|\mathbf{d}|\) is the electric dipole moment of the transition, \(\omega_0\) is the transition angular frequency, and \(\epsilon_0\) is the vacuum permittivity. The total population inversion density is \(n=2n'\), while \(P^+\) and \(E^+\) are the slowly varying amplitudes of the induced polarization and electric field, respectively.
The term \(\Lambda_N\) represents the effective inversion pump used in the main text. The term \(\Lambda_P\) represents a phenomenological polarization source or initial polarization fluctuation used to seed the evolution of the system. The non-coherent relaxation and dephasing processes are described by the terms \(-n'/T_1\) and \(-P^+/T_2\), respectively.

The corresponding polarization and electric-field vectors are written as
\begin{align}
    \mathbf{P}^{\pm}(z,\tau)
    &=
    P^{\pm}(z,\tau)
    e^{\pm i\omega_0\tau}
    \boldsymbol{\epsilon}_d ,
    \label{eq:svea_polarization}
    \\
    \mathbf{E}^{\pm}(z,\tau)
    &=
    E^{\pm}(z,\tau)
    e^{\mp i\omega_0\tau}
    \boldsymbol{\epsilon}_d ,
    \label{eq:svea_field}
\end{align}
where \(\boldsymbol{\epsilon}_d=\mathbf{d}/d\) is the unit vector along the transition dipole moment. The superscripts \(+\) and \(-\) denote positive- and negative-frequency components, with \(P^-=(P^+)^*\) and \(E^-=(E^+)^*\).

%%%%%%%%%%%%%%%%%%%%%%%%%%%%%%%%%%%%%%%%%%%%%%%%%%%%%%%%%%%%%%%%%%%%%%%

\section{Fitted pump parameters and example of multi-component fitting}
\label{app:fit_parameters}

\begin{figure*}[htp!]
    \centering
    \includegraphics[width=0.5\linewidth]{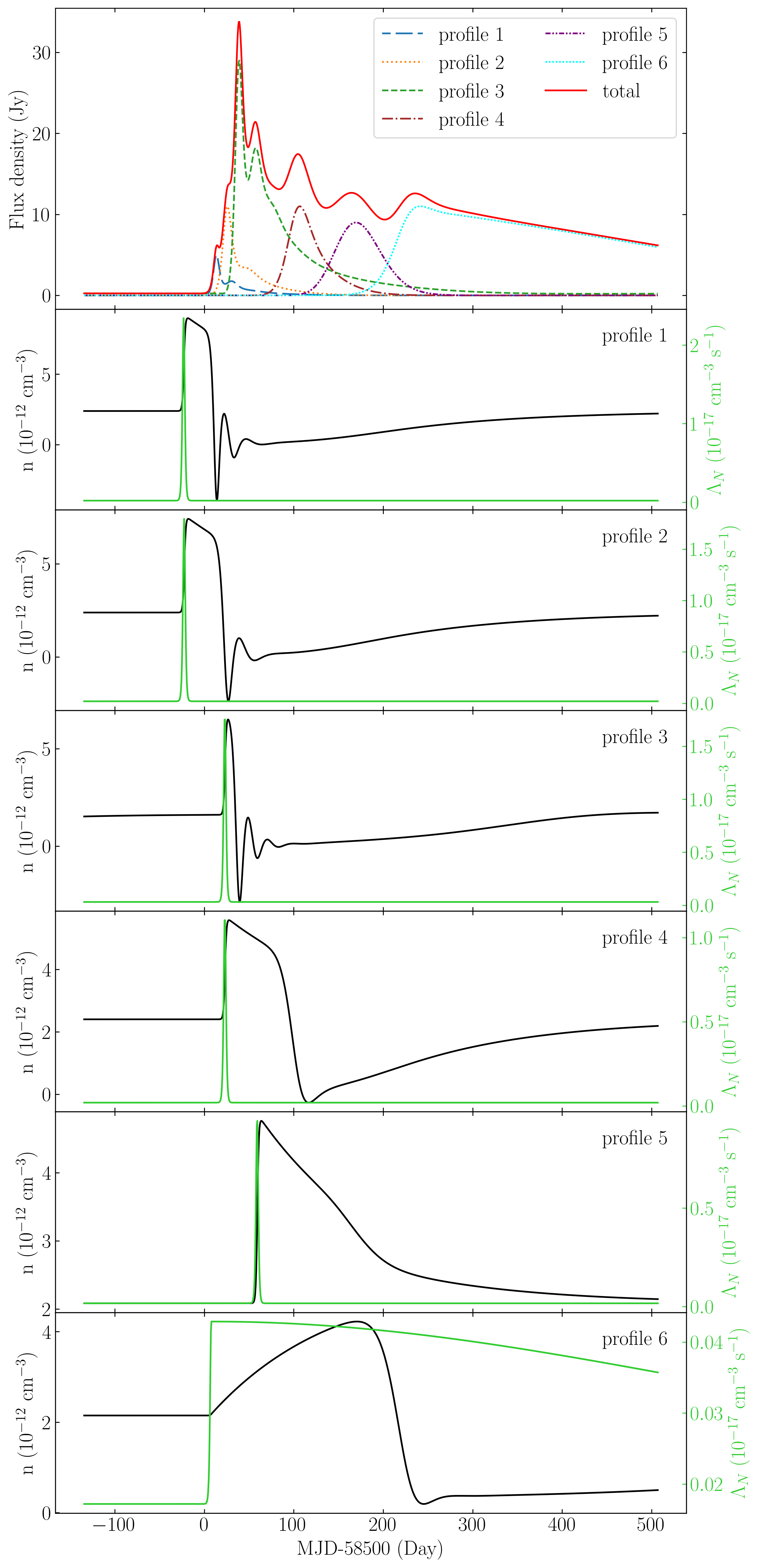}
    \caption{Representative multi-component model for
\(v_{\rm LSR}=-18.05~\kms\). The top panel shows the individual scaled Maxwell--Bloch intensity responses and their summed contribution; the corresponding fit to the iMet light curve is shown in the top panel of Fig.~\ref{fig:v~-18}. The lower panels show, for each fitted profile, the effective pump excitation \(\Lambda_{N,j}\) (green curves, right axes) and the resulting population inversion \(n_j\) (black curves, left axes). Profiles 1--5 use the symmetric pump pulse given by Eq.~\eqref{eq:pump}, whereas profile 6 uses the asymmetric long-tail pump given by Eq.~\eqref{eq:pump_tail}.}\label{fig:m1805_n_lambda}
\end{figure*}

The fitted pump parameters and initial inverted column densities for all
modelled velocity channels are listed in Table~\ref{tab:fit_parameters}.
As a representative example of the multi-component decomposition,
Fig.~\ref{fig:m1805_n_lambda} shows the fit for \(v_{\rm LSR}=-18.05~\kms\). The top panel shows the individual scaled Maxwell--Bloch intensity responses and their summed contribution, which corresponds to the combined model shown in Fig.~\ref{fig:v~-18} (top panel).
The lower panels show, for each fitted profile, the effective pump excitation
and the resulting population inversion response. Notably, the inversion response to the broad pump component of profile 6 (bottom panel) shows an initial superradiance transient before settling into a saturated maser regime \citep{Rajabi2020}. This figure illustrates how
the phenomenological pump histories generate the individual intensity
components used in the fit; the decomposition should be interpreted as representative rather than unique.

\begin{table*}[t!]
\centering
\caption{Fitted pump parameters for the effective Maxwell--Bloch components shown in Figs.~\ref{fig:v~-18}--\ref{fig:v~-15.6}.}
\label{tab:fit_parameters}

\scriptsize
\renewcommand{\arraystretch}{1.18}
\setlength{\tabcolsep}{6pt}

\begin{tabular}{|c|c|c|c|c|c|}
\hline
\(v_{\rm LSR}~(\kms)\) &
Profile &
\(\Lambda_0~(\mathrm{cm^{-3}\,s^{-1}})\) &
\(\Lambda_1~(\mathrm{cm^{-3}\,s^{-1}})\) &
\(\tau_0\)~(d) &
\(n_0L~(\mathrm{cm^{-2}})\) \\
\hline

\multirow{6}{*}{\(-18.05\)}
& 1 & \(1.91\times10^{-19}\) & \(2.33\times10^{-17}\) & 477.8 & \(5.98\times10^3\)\\
& 2 & \(1.91\times10^{-19}\) & \(1.78\times10^{-17}\) & 478.3 & \(5.98\times10^3\)\\
& 3 & \(3.10\times10^{-19}\) & \(1.78\times10^{-17}\) & 523.8 & \(9.71\times10^3\)\\
& 4 & \(1.92\times10^{-19}\) & \(1.12\times10^{-17}\) & 523.8 & \(6.01\times10^3\)\\
& 5 & \(1.67\times10^{-19}\) & \(9.45\times10^{-18}\) & 559.9 & \(5.23\times10^3\)\\
& 6 & \(1.72\times10^{-19}\) & \(2.57\times10^{-19}\) & 508.6 & \(5.39\times10^3\)\\
\hline

\multirow{6}{*}{\(-17.95\)}
& 1 & \(7.16\times10^{-20}\) & \(2.99\times10^{-17}\) & 472.0 & \(2.24\times10^3\)\\
& 2 & \(7.16\times10^{-20}\) & \(2.50\times10^{-17}\) & 473.0 & \(2.24\times10^3\)\\
& 3 & \(3.07\times10^{-19}\) & \(1.55\times10^{-17}\) & 522.2 & \(9.61\times10^3\)\\
& 4 & \(1.67\times10^{-19}\) & \(1.20\times10^{-17}\) & 509.7 & \(5.23\times10^3\)\\
& 5 & \(1.67\times10^{-19}\) & \(1.00\times10^{-17}\) & 562.5 & \(5.23\times10^3\)\\
& 6 & \(1.84\times10^{-19}\) & \(2.04\times10^{-19}\) & 483.5 & \(5.76\times10^3\)\\
\hline

\multirow{5}{*}{\(-17.85\)}
& 1 & \(1.67\times10^{-19}\) & \(1.63\times10^{-17}\) & 464.7 & \(5.23\times10^3\)\\
& 2 & \(1.92\times10^{-19}\) & \(1.59\times10^{-17}\) & 488.2 & \(6.01\times10^3\)\\
& 3 & \(1.92\times10^{-19}\) & \(8.90\times10^{-18}\) & 490.3 & \(6.01\times10^3\)\\
& 4 & \(1.47\times10^{-19}\) & \(9.63\times10^{-18}\) & 566.7 & \(4.60\times10^3\)\\
& 5 & \(1.94\times10^{-19}\) & \(2.04\times10^{-19}\) & 501.3 & \(6.08\times10^3\)\\
\hline

\multirow{5}{*}{\(-15.75\)}
& 1 & \(1.67\times10^{-19}\) & \(1.49\times10^{-17}\) & 462.1 & \(5.23\times10^3\)\\
& 2 & \(1.67\times10^{-19}\) & \(1.41\times10^{-17}\) & 484.0 & \(5.23\times10^3\)\\
& 3 & \(1.67\times10^{-19}\) & \(1.14\times10^{-17}\) & 500.3 & \(5.23\times10^3\)\\
& 4 & \(1.67\times10^{-19}\) & \(1.30\times10^{-17}\) & 563.1 & \(5.23\times10^3\)\\
& 5 & \(3.12\times10^{-19}\) & \(2.48\times10^{-18}\) & 697.0 & \(9.77\times10^3\)\\
\hline

\multirow{5}{*}{\(-15.65\)}
& 1 & \(1.67\times10^{-19}\) & \(1.51\times10^{-17}\) & 463.6 & \(5.23\times10^3\)\\
& 2 & \(1.67\times10^{-19}\) & \(1.43\times10^{-17}\) & 484.6 & \(5.23\times10^3\)\\
& 3 & \(1.14\times10^{-19}\) & \(1.45\times10^{-17}\) & 487.7 & \(3.57\times10^3\)\\
& 4 & \(1.49\times10^{-19}\) & \(1.45\times10^{-17}\) & 565.1 & \(4.67\times10^3\)\\
& 5 & \(1.27\times10^{-19}\) & \(9.88\times10^{-19}\) & 690.2 & \(3.98\times10^3\)\\
\hline

\multirow{5}{*}{\(-15.53\)}
& 1 & \(1.67\times10^{-19}\) & \(1.50\times10^{-17}\) & 463.6 & \(5.23\times10^3\)\\
& 2 & \(1.58\times10^{-19}\) & \(1.58\times10^{-17}\) & 490.8 & \(4.95\times10^3\)\\
& 3 & \(1.13\times10^{-19}\) & \(1.50\times10^{-17}\) & 490.3 & \(3.54\times10^3\)\\
& 4 & \(1.38\times10^{-19}\) & \(1.59\times10^{-17}\) & 571.4 & \(4.32\times10^3\)\\
& 5 & \(2.89\times10^{-19}\) & \(1.20\times10^{-18}\) & 647.3 & \(9.05\times10^3\)\\
\hline
\end{tabular}

\vspace{0.6em}
\begin{minipage}{0.95\textwidth}
\footnotesize
\emph{Notes.} The listed pump rates are the cgs equivalents of the SI values given in the figure captions, using \(1~\mathrm{m^{-3}}=10^{-6}~\mathrm{cm^{-3}}\). The initial inverted column density is calculated as \(n_0L=(\Lambda_0T_1)L\), with \(T_1=144.4\)~d and \(L=2.51\times10^{15}\)~cm. As explained in the discussion surrounding equation~(\ref{eq:pump}), \(\Lambda_0\) and \(\Lambda_1\) denote the constant background pump rate and the pump-pulse amplitude, respectively, while \(\tau_0\) denotes the arrival time of the excitation and is given in days. The profiles shown in Figs.~\ref{fig:v~-18}--\ref{fig:v~-15.6} can be reproduced using these values, together with \(T_2=6.9\)~d and a pump-pulse duration of \(T_\mathrm{p}=1.7\)~d.
\end{minipage}

% \begin{minipage}{0.95\textwidth}
% \footnotesize
% \emph{Notes.} The listed pump rates are the cgs
% equivalents of the SI values used in the figure captions, with
% \(1~\mathrm{m^{-3}}=10^{-6}~\mathrm{cm^{-3}}\). The initial inverted column
% density is calculated from \(n_0L=(\Lambda_0T_{1})L\), with
% \(T_1=144.4\)~d and \(L=2.51\times10^{15}\)~cm. As mentioned in the discussion surround equation (\ref{eq:pump}), $\Lambda_0$ and $\Lambda_1$ are the constant background pump and pump pulse amplitude, respectively. The arrival time of the excitation is denoted by $\tau_0$. With $T_2=6.9$~d and the pulse width $T_\mathrm{p}=1.7$~d. The profiles in Fig.~\ref{fig:v~-18}--\ref{fig:v~-15.6} can be reproduced using these values.
% \end{minipage}

\end{table*}

%% For example:

%% \bibitem[Aladro et al.(2015)]{Aladro15} Aladro, R., Martín, S., Riquelme, D., et al. 2015, \aas, 579, A101

%%\end{thebibliography}

\end{document}